\documentclass[twocolumn]{aastex63}
\usepackage{graphicx}
\usepackage{color}
\usepackage{amsmath,amssymb}
\usepackage{gensymb}
\usepackage{times}
\usepackage{nicefrac}
\usepackage{xcolor}

\usepackage[utf8]{inputenc}
\begin{document}
\title{The observable properties of galaxy accretion events in Milky Way-like galaxies in the FIRE-2 cosmological simulations}

\author[0000-0003-1856-2151]{Danny Horta}
\affiliation{Center for Computational Astrophysics, Flatiron Institute, 162 5th Ave., New York, NY 10010, USA\\}
\affiliation{Astrophysics Research Institute, Brownlow Hill,  Liverpool, L3 5RF, UK\\}
\author[0000-0002-6993-0826]{Emily C. Cunningham}
\altaffiliation{NASA Hubble Fellow}
\affiliation{Center for Computational Astrophysics, Flatiron Institute, 162 5th Ave., New York, NY 10010, USA\\}
\affiliation{Department of Astronomy, Columbia University, 550 West 120th Street, New York, NY, 10027, USA\\}
\author[0000-0003-3939-3297]{Robyn E. Sanderson}
\affiliation{Center for Computational Astrophysics, Flatiron Institute, 162 5th Ave., New York, NY 10010, USA\\}
\affil{Department of Physics \& Astronomy, University of Pennsylvania, 209 S 33rd St., Philadelphia, PA 19104, USA\\}
\author{Kathryn V. Johnston}
\affiliation{Center for Computational Astrophysics, Flatiron Institute, 162 5th Ave., New York, NY 10010, USA\\}
\affiliation{Department of Astronomy, Columbia University, 550 West 120th Street, New York, NY, 10027, USA\\}
\author[0000-0001-5214-8822]{Nondh Panithanpaisal}
\affil{Department of Physics \& Astronomy, University of Pennsylvania, 209 S 33rd St., Philadelphia, PA 19104, USA\\}
\author[0000-0002-8354-7356]{Arpit Arora}
\affil{Department of Physics \& Astronomy, University of Pennsylvania, 209 S 33rd St., Philadelphia, PA 19104, USA\\}
\author{Lina Necib}
\affil{The NSF AI Institute for Artificial Intelligence and Fundamental Interactions}
\affil{Department of Physics and Kavli Institute for Astrophysics and Space Research, Massachusetts Institute of Technology, 77 Massachusetts Ave, Cambridge MA 02139, USA\\}
\author[0000-0003-0603-8942]{Andrew Wetzel}
\affil{Department of Physics and Astronomy, University of California, Davis, CA 95616, USA}
\author{Jeremy Bailin}
\affil{Department of Physics and Astronomy, University of Alabama, Tuscaloosa, AL, 35487, USA}
\author{ Claude-Andr\'e Faucher-Gigu\`ere}
\affil{Department of Physics and Astronomy and CIERA, Northwestern University, 1800 Sherman Avenue, Evanston, IL 60201, USA}
\correspondingauthor{Danny Horta}
\email{dhortadarrington@flatironinstitute.org}

\begin{abstract}
In the $\Lambda$-Cold Dark Matter model of the Universe, galaxies form in part through accreting satellite systems. Previous work have built an understanding of the signatures of these processes contained within galactic stellar halos. This work revisits that picture using seven Milky Way-like galaxies in the \textit{Latte} suite of FIRE-2 cosmological simulations. The resolution of these simulations allows a comparison of contributions from satellites above M$_{*}$$\gtrsim$10$\times$$^{7}$M$_{\odot}$, enabling the analysis of observable properties for disrupted satellites in a fully self-consistent and cosmological context. Our results show that, the time of accretion and the stellar mass of an accreted satellite are fundamental parameters that in partnership dictate the resulting spatial distribution, orbital energy, and [$\alpha$/Fe]-[Fe/H] compositions of the stellar debris of such mergers \textit{at present day}. These parameters also govern the resulting dynamical state of an accreted galaxy at $z=0$, leading to the expectation that the inner regions of the stellar halo (R$_{\mathrm{GC}}$ $\lesssim$30 kpc) should contain fully phase-mixed debris from both lower and higher mass satellites. In addition, we find that a significant fraction of the lower mass satellites accreted at early times deposit debris in the outer halo (R$_{\mathrm{GC}}$ $>$50 kpc) that are {\it not} fully phased-mixed, indicating that they could be identified in kinematic surveys. Our results suggest that, as future surveys become increasingly able to map the outer halo of our Galaxy, they may reveal the remnants of long-dead dwarf galaxies whose counterparts are too faint to be seen {\it in situ} in higher redshift surveys.
\end{abstract}


\section{Introduction}
\label{Introduction}
A fundamental prediction from the current cosmological paradigm --$\Lambda$-CDM-- is that galaxies grow in great measure via the process of hierarchical mass assembly (\citealp[][]{Searle1978,Bullock2005}). As a result, stellar halos of galaxies are expected to host large quantities of accreted material in the form of phase-mixed debris, streams, and satellite galaxies (\citealp[e.g.,][]{Johnston1996,Helmi1999,Bullock2005,Johnston2008,Cooper2010,Font2011}). This accreted component is unique for each galaxy, and is dictated by its accretion history (\citealp[e.g.,][]{Bullock2005, Johnston2008,DeLucia2008,Gomez2012,Tissera2012}). 

Halo substructures believed to be the remains of cannibalised systems have been discovered in the Milky Way (\citealp[MW; e.g.,][]{Ibata1994,Helmi1999,Belokurov2018,Helmi2018,Haywood2018,Mackereth2019,Myeong2019,Koppelman2019b,Naidu2020,Yuan2020,Horta2021}), Andromeda (\citealp[e.g.,][]{Ibata2007,McConnachie2009,Veljanoski2014}), and nearby galaxies (\citealp[e.g.,][]{Martinez2008,Atkinson2013,Duc2015,Crnojevi2016,Chiti2021}). The characterization of such debris in observable spaces such as kinematics, chemistry, and age provide an opportunity to study the remains of the many galaxies destroyed by the MW that did not survive to $z=0$ on a star by star basis, and enable the opportunity to develop a deeper understanding on how these building blocks contribute to the formation and evolution of the Galaxy.

We are now in the era of large-scale stellar surveys of the MW. The data collected from the revolutionary \textit{Gaia} mission \citep[][]{Gaia2018,Gaiaedr32021}, as well as the latest wide-field spectroscopic programs such as APOGEE \citep[][]{Majewski2017}, GALAH \citep[][]{DeSilva2015}, RAVE \citep[][]{Steinmetz2020}, LAMOST \citep[][]{Cui2012}, and H3 \citep[][]{Conroy2019}, amongst others, are providing a colossal archive of chemo-dynamical properties for millions of stars in the MW. This is paving the way for new discoveries of substructure in the stellar halo component of the Galaxy (and halos of other Local Group galaxies), as well as tighter constraints on the density profile and shape (\citealp[e.g.,][]{Deason2011,Iorio2018,Iorio2019,Deason2019, Deason2021}), the mass of its different stellar populations (\citealp[e.g.,][]{Mackereth2020,Deason2021, Horta2021_nrich}), its chemical properties (\citealp[e.g.,][]{Schiavon2017_nrich,Hayes2018,Mackereth2019,Matsuno2019,Monty2020,Buder2021, Kisku2021,Horta2022}), and the perturbations to its stellar/dark matter components from its interaction with its massive satellites (\citealp[e.g.,][]{Cunningham2020, Erkal2020, Garavito2020, Conroy2021, Vasiliev2021}). However, although the great strides performed on the observational front are shedding light on the nature and accretion history of our Galaxy, there is still a demand to interpret such observational findings theoretically in a cosmological setting using high-resolution simulations.

Since the classic papers by \citet{Bullock2005} and \citet{Johnston2008}, many works have sought to interpret the resulting phase-space and orbital distribution of stars in the stellar halos of galaxies from a theoretical standpoint (\citealp[e.g.,][]{Cooper2010,Pillepich2014, Amorisco2017}), and specifically for MW-like galaxies (\citealp[e.g.,][]{Font2011, McCarthy2012, Deason2013, Deason2015, Deason2016, Amorisco2017_mw, Dsouza2018, Monachesi2019, Evans2020, Fattahi2020,Font2020,Grand2020,Santistevan2020,Cunningham2022,Grand2021,Nikakhtar2021, Panithanpaisal2021}). The results from such efforts have helped reveal the diversity and complexity of substructure in the stellar halos of galaxies like our own MW, and have helped shed light on the properties of discovered debris in the stellar halo of the Galaxy resulting from hierarchical mass assembly. However, many of these results used in: $i$) the regime of tailored/idealized $N$-body simulations; $ii$) the context of large cosmological simulations that do not match the high-resolution needed to study the intricacies of this hierarchical formation process.

In this work, we set to test if expectations from tailored $N$-body simulations survive when examining the observable properties of the debris from disrupted satellite galaxies using high-resolution (zoom-in) cosmological simulations. We ask three main questions: 

\begin{enumerate}
    \item ``\textit{What are the  properties of disrupted satellites?}"
    \item ``\textit{Do our expectations from prior simulations survive in a high-resolution and cosmological setting}?"
    \item ``\textit{What do these results mean for future surveys}"?
\end{enumerate}

We make use of the sophisticated Feedback In Realistic Environments (\citealp[i.e., FIRE-2,][]{Hopkins2018}) suite of numerical cosmological simulations to explore the distributions of halo star properties in spatial, orbital, and chemical planes as a function of their progenitor stellar mass and infall time. Our aim is to understand the relationship between merger mass and infall time with the distribution of the resulting stellar particles of independent merger events in diagnostic planes that can be compared with observational findings. Moreover, we also aim to obtain results that can be used as a theoretical blueprint for upcoming surveys, such as WEAVE \citep[][]{Dalton2014}, 4MOST \citep[][]{DeJong2019}, DESI \citep[][]{Allende2020}, MWM/SDSS-V \citep[][]{Kollmeier2017}, or the Rubin Observatory's LSST \citep[]{Ivezic2019}, which will likely pave the way for new discoveries of debris from accreted satellites lurking in the outer regions of the Galactic stellar halo, and the characterization of the ones already known.

This paper is structured as follows: in Section~\ref{sims} we describe the simulations used in this work and the methodology employed to identify accretion events in the \textit{Latte}/FIRE-2 cosmological simulations. Subsequently, we present the results obtained in this work, where we show how different mass/infall time accretion events deposit the bulk of their stars as a function of their spatial coordinates in Section~\ref{spatial_distribution}, their orbital parameters in Section~\ref{kinematics}, and their chemical compositions in Section~\ref{Chemistry}. We then discuss the implications of our results and present our concluding statements in Section~\ref{conclusion}.

\section{Simulations and satellite sample} 
\label{sims}

\subsection{Simulations}

We make use of seven MW-like galaxies from the \textit{Latte} suite of FIRE-2\footnote{FIRE project website: \url{http://fire.northwestern.edu}} cosmological zoom-in simulations \citep{Wetzel2016}. These simulations were run with the FIRE-2 physics model \citep{Hopkins2018}, utilising the Lagrangian meshless finite-mass $N$-body gravitational plus hydrodynamics code \texttt{GIZMO}\footnote{http://www.tapir.caltech.edu/~phopkins/Site/GIZMO.html} \citep{Hopkins2015}. FIRE-2 simulations model several radiative cooling and heating processes for gas such as free-free emission, photoionization/recombination, Compton scattering, photoelectric, metal-line, molecular, fine-structure, dust-collisional, and cosmic-ray heating across a temperature range of 10 - 10$^{10}$K. These simulations also include the spatially redshift-dependent and spatially uniform cosmic UV background from \citet{Faucher2009}, for which HI reionization occurs at $z_{\mathrm{reion}}$ $\sim$ 10. Moreover, FIRE-2 self-consistently generates and tracks 11 chemical abundance species (namely,  H, He, C, N, O, Ne, Mg, Si, S, Ca, and Fe), including sub-grid diffusion of these abundances in gas via turbulence (\citealp{Hopkins2016,Su2017,Escala2018}), as well as enrichment from stellar winds.

Star formation occurs in gas that is self-gravitating, Jeans-unstable, cold (T $<$ 10$^{4}$K), dense ($n$ $>$ 1,000 cm$^{-3}$), and molecular (following \citealp{Krumholz2011}). As a star particle forms it inherits the mass and chemical abundance composition from its progenitor gas cell, and represents a single stellar population with particle mass of 7,100 M$\odot$, assuming a \citet{Kroupa2001} initial mass function. Star particles then evolve along stellar population models from \texttt{STARBURST} v7.0 \citep{Leitherer1999}. FIRE-2 simulations include several different feedback processes, including core-collapse and Ia supernovae, mass loss from stellar winds, and radiation, including radiation pressure, photoionization, and photo-electric heating. 

The \textit{Latte} suite was generated within periodic cosmological boxes of lengths 70.4 - 172 Mpc using the code \texttt{MUSIC} \citep{hahn2011}, and employing cosmological zoom-in initial conditions for each simulation at $z$ $\simeq$ 99. Each simulation has 600 snapshots saved down to $z$ = 0, spaced every $\simeq$ 25 Myr. All simulations assume flat $\Lambda$-CDM cosmology with parameters consistent with \citet{Plank2020}. More specifically, the \textit{Latte} suite (excluding m12r and m12w) used $\Omega_{\mathrm{m}}$ = 0.272, $\Omega_{\mathrm{b}}$ = 0.0455,
$\sigma_{\mathrm{8}}$ = 0.807, $n_{\mathrm{s}}$ = 0.961, $h$ = 0.702. The m12r and m12w halos were selected specifically because they host an LMC-mass satellite galaxy, and adopted updated initial conditions from \citep{Plank2020} compared to their MW-mass galaxy siblings: $h$ = 0.68, $\Omega_{\mathrm{\Lambda}}$ = 0.31, $\Omega_{\mathrm{m}}$ = 0.31, $\Omega_{\mathrm{b}}$ = 0.048,
$\sigma_{\mathrm{8}}$ = 0.82, $n_{\mathrm{s}}$ = 0.961, $h$ = 0.97.
 
The resolution of this suite of simulations enables baryonic subhalos to be well resolved even near each MW-like galaxy. It also resolves the formation of tidal streams from satellite galaxies down to approximately 10$^{8}$M$_{\odot}$ in total mass or 10$^{6}$M$_{\odot}$ in stellar mass (at $z$=0), similar to that of the MW's ``classical" dwarf spheroidals (\citealp[e.g.,][]{Panithanpaisal2021,Cunningham2022,Shipp2022}). Furthermore, dark matter particles in each snapshot are processed with \texttt{Rockstar} \citep[][]{Behroozi2013} to produce halo catalogs that are connected in time using \texttt{consistent trees} to form a merger tree \citep[][]{Behroozi2013b}. Once a merger tree is constructed, a preliminary assignment of star particles to each halo and subhalo, identified by \texttt{Rockstar} in each snapshot, is made by selecting star particles within the halo's virial radius and within twice the halo circular velocity relative to the halo's centre. Thus, all galaxies are tracked within the simulated volume across all snapshots. The post-processing is done using \texttt{gizmo analysis} \citep[][]{Wetzel2020a} and \texttt{halo analysis} \citep[][]{Wetzel2020b}.

The  properties of the host galaxies in \textit{Latte} show broad agreement with the MW, including the stellar-to-halo mass relation \citep[][]{Hopkins2018}, stellar halos (\citealp[][]{Bonaca2017, Sanderson2018}), and the radial and vertical structure of their disks (\citealp[][]{Ma2017,Bellardini2021}). Moreover, the satellite populations of these suite of simulations have also been demonstrated to agree with several observed properties, such as: the mass and velocity dispersions (\citealp[][]{Wetzel2016,Garrison2019_dispersions}); star formation histories \citep[][]{Garrison2019_mass}; and radial distributions \citep[][]{Samuel2020}. Despite great similarities in the properties of the satellite galaxies in \textit{Latte} and that of the observed satellites around the MW, it has been shown that the former are generally too metal-poor when compared to the latter (\citealp[][]{Escala2018,Wheeler2019,Panithanpaisal2021}). This is likely a consequence of the delay time distributions of SNIa \citep{Gandhi2022}. For this work, we emphasize that we do not require quantitative agreement with simulated and observed abundances. The relations found in this paper between (luminous) subhalo parameters and their respective chemical abundances should be treated qualitatively, and are intended for use within the simulations only. However, although the \emph{normalization} of the various abundances is not always in good agreement with observations, we expect the \emph{trends} we identify in this paper to be robust.

\subsection{Identifying accretion events in the simulations}
\label{mergers}

In this subsection we describe the method for identifying accretion events in each cosmological simulation. Each accretion event identified in this work was selected based on a set of criteria, building on the selection method from \citet{Panithanpaisal2021}. 

\begin{figure*}
\centering
\includegraphics[width=\textwidth]{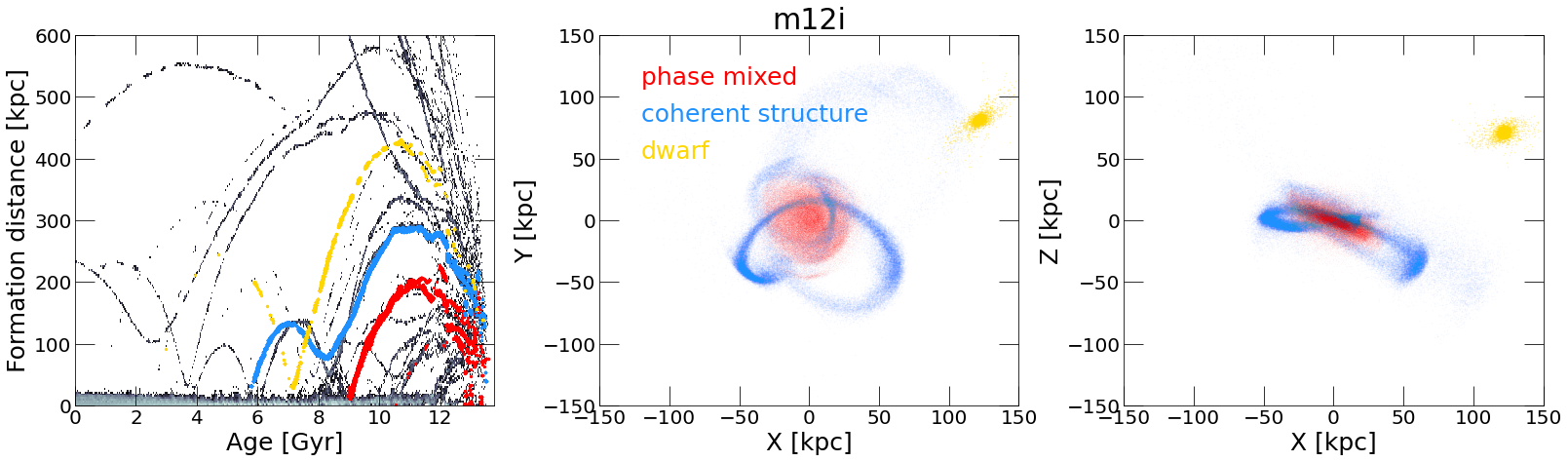}
\caption{\textit{Left}: Density distribution showing the formation distance as a function of age for all the star particles in the \textit{Latte} halo \texttt{m12i}. Here every "branch"-like feature corresponds to a different subhalo, some of which subsequently merge with the main host (MW-like) halo with time (see \citet[][]{Cunningham2022} for an introduction to this diagnostic plane). Highlighted as red/blue/yellow branches are three accretion events identified for this simulation, corresponding to a phase mixed/coherent stream/dwarf classified accretion events, respectively. \textit{Middle}: face-on projection of the star particles associated with the illustrated accretion events centred on the \texttt{m12i} MW-like Galaxy. \textit{Right}: star particles associated with the illustrated accretion events, again centred on the \texttt{m12i} MW-like Galaxy, but now from an edge-on projection.}
    \label{formationd_age}
\end{figure*}

The first step involves inspecting the density distribution of star particles in each simulation in the age--formation distance plane at $z=0$, first introduced in \citet[][]{Cunningham2022} and \citet[][]{Khoperskov2022_a}, as shown in Fig\ref{formationd_age}. Here, star particle ages are shown on the $x$-axis, and the formation distance of star particles (defined as the distance of the star particle at its formation time from the center of the main MW-like host) is on the $y$-axis. The grayscale color in this panel indicates the density of star particles for the \textit{Latte} galaxy \texttt{m12i} (first introduced in \citet[][]{Wetzel2016}\footnote{The seven MW-like galaxies in $Latte$ studied in this work are referred to as \texttt{m12b}, \texttt{m12c}, \texttt{m12f}, \texttt{m12i}, \texttt{m12m}, \texttt{m12r}, \texttt{m12w}}). In this plane, many branch structures (i.e., luminous subhalos) appear to merge with the main MW host of the simulation (namely, the high density region at formation distance $<$25 kpc). Star particles that form in the branches at large formation distances are stars forming in dwarf galaxies, some of which enter the main host halo and are disrupted. Thus, we use the branches in this plane to identify satellite galaxies accreted by the main host galaxy.

Once a clear merger event was identified in the age-formation distance plane, the next step involved using the halo catalogue for each simulation to identify the luminous subhalo ID that had the smallest distance to the MW-like host at the time in which the subhalo branch in Fig~\ref{formationd_age} connected with the host MW-like galaxy branch. By identifying the subhalo ID with the smallest distance, we were able to identify the star particles belonging to the subhalo of interest, using the particle tracking function built within \texttt{gizmo analysis}. We then tracked star particles for that subhalo to $z$=0 and to the point in which the subhalo first crosses the virial radius of the MW-like host (that is, at the first snapshot in the simulation after the luminous subhalo crosses the virial radius of the host).

 The final sample of disrupted satellites used in this work is comprised of 62 accretion events with stellar masses greater than M$_{\star}$ $>$10$^{7}$M$_{\odot}$. We choose to only study disrupted satellites above this mass threshold as we aim to characterise the observable properties of the debris from the largest (stellar) mass building blocks of the $Latte$ MW-like galaxies. Of this sample of disrupted satellites, 9 belong to m12b, 6 to m12c, 10 to m12f, 12 to m12i, 14 to m12m, 4 to m12r, and 7 to m12w. From our sample, 35 disrupted satellites have a stellar mass smaller than M$_{\star}$ $<$ 10$^{8}$M$_{\odot}$, 19 have a stellar mass between 10$^{8}$M$_{\odot}$ $<$ M$_{\star}$ $<$ 10$^{9}$M$_{\odot}$, and 8 have a stellar mass greater than M$_{\star}$ $>$ 10$^{9}$M$_{\odot}$. For more details on the properties of these identified disrupted satellites, we refer the reader to Table~\ref{merger_summary}.

As \citet{Panithanpaisal2021} recently studied stream progenitors in the suite of simulated MW-like galaxies we use in this work, it is worth comparing the overlap between the corresponding samples. We use a higher mass cut of M$_{\star}$ $>$ 10$^{7}$ M$_{\odot}$ in this work, rather than the M$_{\star}$ = 10$^{5.5}$ M$\odot$ used in \citet{Panithanpaisal2021}. Moreover, \citet[][]{Panithanpaisal2021} selected stream candidates requiring that they be bound $\sim$ 2.7--6.5 Gyr ago. Therefore, many of the earlier mergers identified in this work are not included in their catalogue. \citet[][]{Panithanpaisal2021} also imposed an upper cutoff on the number of star particles in order to rule out objects that are more massive than the Sagittarius stream, which we did not apply in this work. In summary, we find that 22 of the (massive) accretion events identified by \citet{Panithanpaisal2021} are contained within the initial sample of accretion events identified in this work.

\begin{deluxetable*}{cccccl}
\centering
\tablenum{1}
\tabletypesize{\small}
\tablewidth{3cm}
\tablecaption{Summary of properties for the massive accretion events identified in this work for every MW-like halo in the \textit{Latte} suite of FIRE-2 simulations. $^{\dagger}$ The mean values quoted are only applicable to accretion events studied in this work. \label{merger_summary}}
\tablehead{
 \colhead{Simulation} &  \colhead{N$_{\mathrm{accretions}}$} &  \colhead{M$_{\mathrm{*, accretion}}$ [M$_{\odot}$]}&  \colhead{$\tau_{\mathrm{infall}}$ [Gyr]} &  \colhead{Classification}\\
 \hline
 \hline
 & & \colhead{(10$^{7}$-10$^{8}$/10$^{8}$-10$^{9}$/$>$10$^{9}$)}& \colhead{($<$4/4-8/$>$8)} & \colhead{(phase mixed/coherent stream/dwarf)}}
\startdata
\hline
m12b & 9& 6--2--1   & 0--2--7 & 5--3--1 \\
\hline
m12c  & 6& 2--2--2  & 3--1--2  & 2--3--1\\
\hline
m12f &11 & 6--3--2 & 0--4--7  & 7--4--0\\
\hline
m12i & 12& 8--4--0 & 0--2--10  & 4--7--1\\
\hline
m12m & 14& 8--5--1  & 1--2--11 & 7--2--4\\
\hline
m12r & 4& 1--1--2  & 3--0--1 & 2--1--1\\
\hline
m12w & 7& 4--2--1  & 0--3--4  & 6--0--1\\
\hline
Total & 63 & 35--19--9 & 7--14--42 & 33--20--9\\
\hline
Mean$^{\dagger}$ & 9 & 5--2.7--1.3 & 1--2--6 & 4.7--2.8--1.3\\
\hline
\hline
\enddata 
\end{deluxetable*}

\begin{figure}
\includegraphics[width=\columnwidth{}{}]{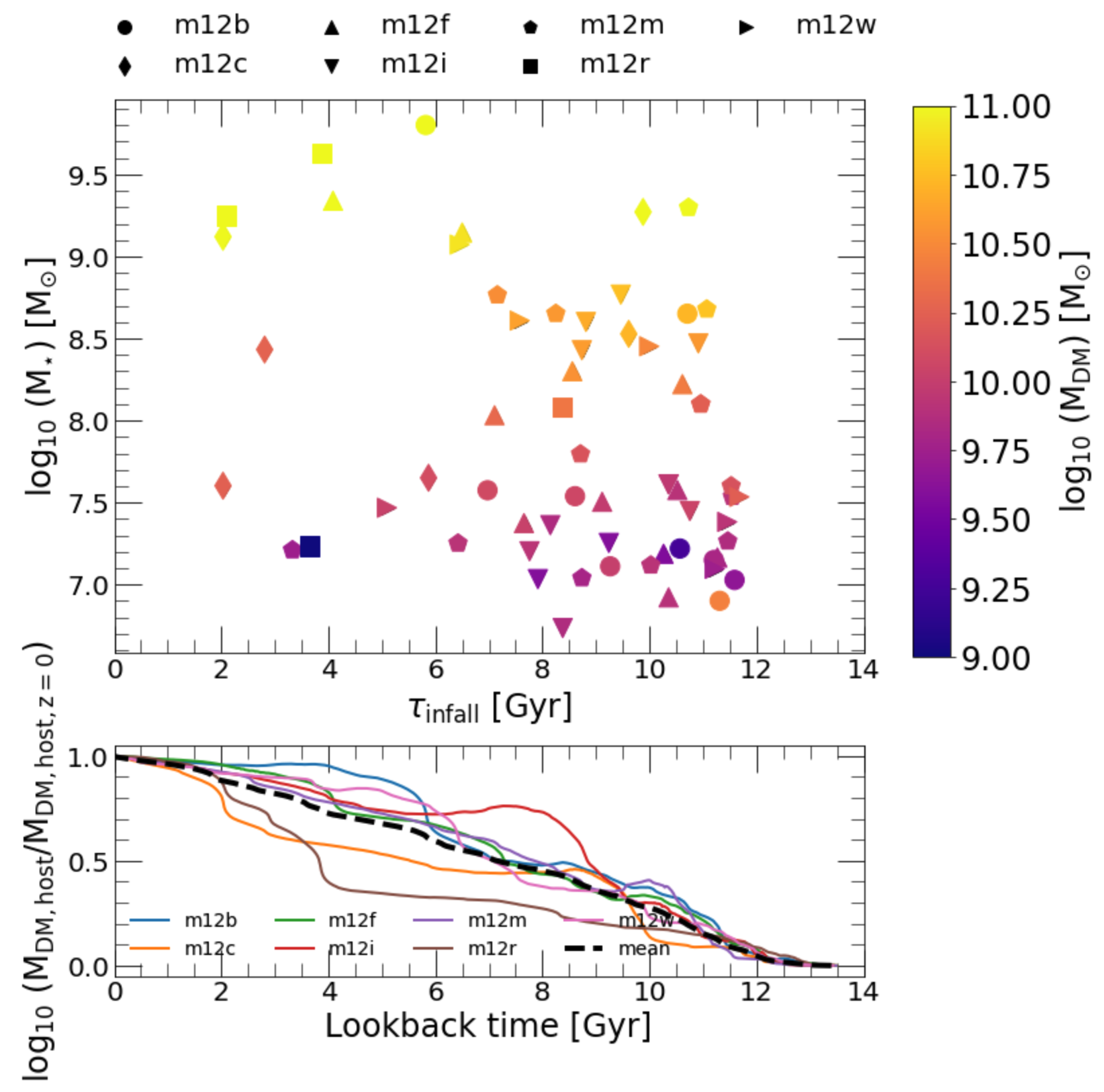}
\caption{Stellar mass at $\tau_{\mathrm{infall}}$ for every accretion event identified in this work as a function of their $\tau_{\mathrm{infall}}$ and color coded by their virial mass (also at $\tau_{\mathrm{infall}}$). The bottom panel shows the growth of each MW-like galaxy as a function of its final virial mass,
where we show the mean as a dashed black line. }
    \label{summary_plot}
\end{figure}

Fig~\ref{summary_plot} shows the stellar mass of every luminous subhalo at infall time as a function of infall time (defined as the lookback time when the progenitor subhalo first crosses the virial radius of the host MW-like galaxy) for all the mergers identified in this study, color coded by the virial mass of the progenitor subhalo at infall. The lower panel of Fig~\ref{summary_plot} shows the time evolution of the virial mass (i.e., M$_{\mathrm{DM,host}}$) for the host MW-like galaxies.
Upon determining the final sample of accretion events in the seven MW-like halos studied, we set out to classify the identified mergers based on how their particles are distributed in phase-space at present day. To do so, we follow the method from \citet{Panithanpaisal2021} (see their Fig 2) and classify the identified mergers into one of the following categories: phase mixed debris, coherent stream debris, or bound dwarf/satellite. For a full description of the method used to define the degree of phase mixing (referred to in the rest of this work as dynamical state) of all the identified accretion events we refer the reader to Section 3.2 in \citet{Panithanpaisal2021}. However, we describe the main steps here for clarity and completeness:

\begin{itemize}
    \item Accretion events that have a maximum pairwise separation distance between any two star particles greater than 120 kpc are classified as either coherent streams or phase mixed candidates. Accretion events that did not satisfy this condition were considered dwarfs/satellites (i.e., accretion events that remain clustered in position space).
    \item Accretion events that have a maximum pairwise separation distance between any two star particles greater than 120 kpc were divided into two further subgroups. Following the method in \citet[][]{Panithanpaisal2021}, phase mixed accretions were distinguished from coherent streams based on an average local velocity dispersion threshold (see Eq~\ref{eq_class}). To do so, \citet[][]{Panithanpaisal2021} determines the local velocity dispersion of a star particle  using its 7--20 (depending on the size of the system) nearest neighbours in phase space, and then determines the median over all star particles in the system to determine the average local velocity dispersion. This local velocity dispersion threshold is $\sim$20 km s$^{-1}$ for a stream candidate with M$_{\star}$$\sim$10$^{7}$M$_{\odot}$ (see their Figure 1). The analytical expression used by \citet[][]{Panithanpaisal2021} to classify accreted debris can be written as:

\begin{equation}
\langle \sigma \rangle = -5.28 \log\Big(\frac{M_{\star}}{M_{\odot}}\Big) + 53.55,
\label{eq_class}
\end{equation}

in units of km$^{-1}$. \citet[][]{Panithanpaisal2021} determined this relation by using a linear kernel Support Vector Machine (SVM) with disrupted systems in all of the $Latte$ suite of simulations. In practice, coherent streams have lower average local velocity dispersions than phase mixed debris at fixed M$_{\star}$. A summary of the mergers identified and their corresponding properties are listed in Table~\ref{merger_summary}, while a summary of the dynamical state classification and parameter definitions employed in this work are provided in Table~\ref{definitions}.

\end{itemize}

In the following Sections we present the main results obtained in this study: the resulting distributions of massive accretion events in MW-like Galaxies in spatial, orbital, and chemical planes.

\begin{deluxetable*}{ccl}
\centering
\tablenum{2}
\tabletypesize{\small}
\tablewidth{3cm}
\tablecaption{Useful definitions. \label{definitions}}
\tablehead{
\colhead{} & \colhead{Merger classification}}
\startdata
\hline
\hline
Phase mixed debris & Accretion event with a maximum pairwise separation distance between any two star particles greater than 120 kpc,\\ & whose local velocity dispersion of star particles falls above the threshold given by Equation~\ref{eq_class} \\
\hline
Coherent stream debris & Accretion event with a maximum pairwise separation distance between any two star particles greater than 120 kpc,\\ & whose local velocity dispersion of star particles falls below the threshold given by Equation~\ref{eq_class}\\
\hline
Dwarf/satellite & Accretion event with a maximum pairwise separation distance between any two star particles smaller than 120 kpc\\
\hline
\hline
& Parameter definitions\\
\hline
\hline
$\tau_{\mathrm{infall}}$ [Gyr] & Lookback time in the simulation corresponding to the snapshot after which a subhalo crosses the virial radius of the host \\ & MW-like Galaxy. For this work, $\tau_{\mathrm{infall}}$ is referred to as the time of the accretion event, where the beginning/end of \\ & the simulation corresponds to a $\tau_{\mathrm{infall}}$ value of 14/0 [Gyr], respectively\\
\hline
Formation distance [kpc] & Distance of a simulation star particle from the center of the host MW-like galaxy at the time that the star particle was formed\\
\hline
M$_{\star}$ [M$_{\odot}$] & Stellar mass of the subhalo at $\tau_{\mathrm{infall}}$\\
\hline
M$_{\mathrm{DM}}$ [M$_{\odot}$] & Virial mass of the subhalo at $\tau_{\mathrm{infall}}$\\
\hline
L$_{\mathrm{tot}}$ [kpc kms$^{-1}$] & Total angular momentum per unit mass of a system at $z$=0, defined as L$_{\mathrm{tot}}$ = $\sqrt{\mathrm{L}_{x}^{2}+\mathrm{L}_{y}^{2}+\mathrm{L}_{z}^{2}}$, where [$\mathrm{L}_{x},\mathrm{L}_{y},\mathrm{L}_{z}$] \\ & are the three Cartesian coordinates comprising the angular momentum vector L  centred on the host MW-like \\ & galaxy, and L = $r \times v$. Here $r$/$v$ correspond to the three component Cartesian coordinate position and velocity vector \\ & of every star particle with respect to the host MW-like galaxy coordinate system at $z$=0\\
\hline
E [km$^{2}$s$^{-2}$]& Orbital energy per unit mass of a system at $z$=0, computed as the average of the orbital energy of all star particles in a \\ & system weighted by a particles stellar mass, defined as the sum of the potential and kinetic energies (i.e., E = $\Phi$ + $K$).\\& In order to obtain all the MW-like halo energies on the same scale, these are normalized by the orbital energy at \\& $\langle$R$_{\mathrm{GC}}$ $\rangle$ = 200 kpc (namely, E$_{200}$)\\
\hline
$\eta$ & Orbital circularity at $z$=0, defined as L$_{\mathrm{tot}}$(E)/L$_{\mathrm{circ}}$(E). Here, L$_{\mathrm{circ}}$ equates to the angular momentum of a circular \\ & orbit with the same orbital energy as each star particle in a system, defined as L$_{\mathrm{circ}}$ = $r_{\mathrm{circ}}$ $\times$ $v_{\mathrm{circ}}$ (where $r_{\mathrm{circ}}$ and $v_{\mathrm{circ}}$ \\ & are the position and velocity vectors, respectively, of a star particle centred on the host MW-like galaxy coordinate system \\ & on a circular orbit with the same orbital energy)\\
\hline
\hline
\enddata
\end{deluxetable*}

\section{Spatial distribution}
\label{spatial_distribution}

We begin our analysis with an examination of the spatial distribution of the disrupted satellites. This is a property of the accretion events that should be easiest to compare to observational findings, and which may serve as important blueprints for current/upcoming surveys aiming at characterising the stellar halo of the Milky Way and Andromeda galaxies. 

\begin{figure*}
\centering
\includegraphics[width=\textwidth{}{}]{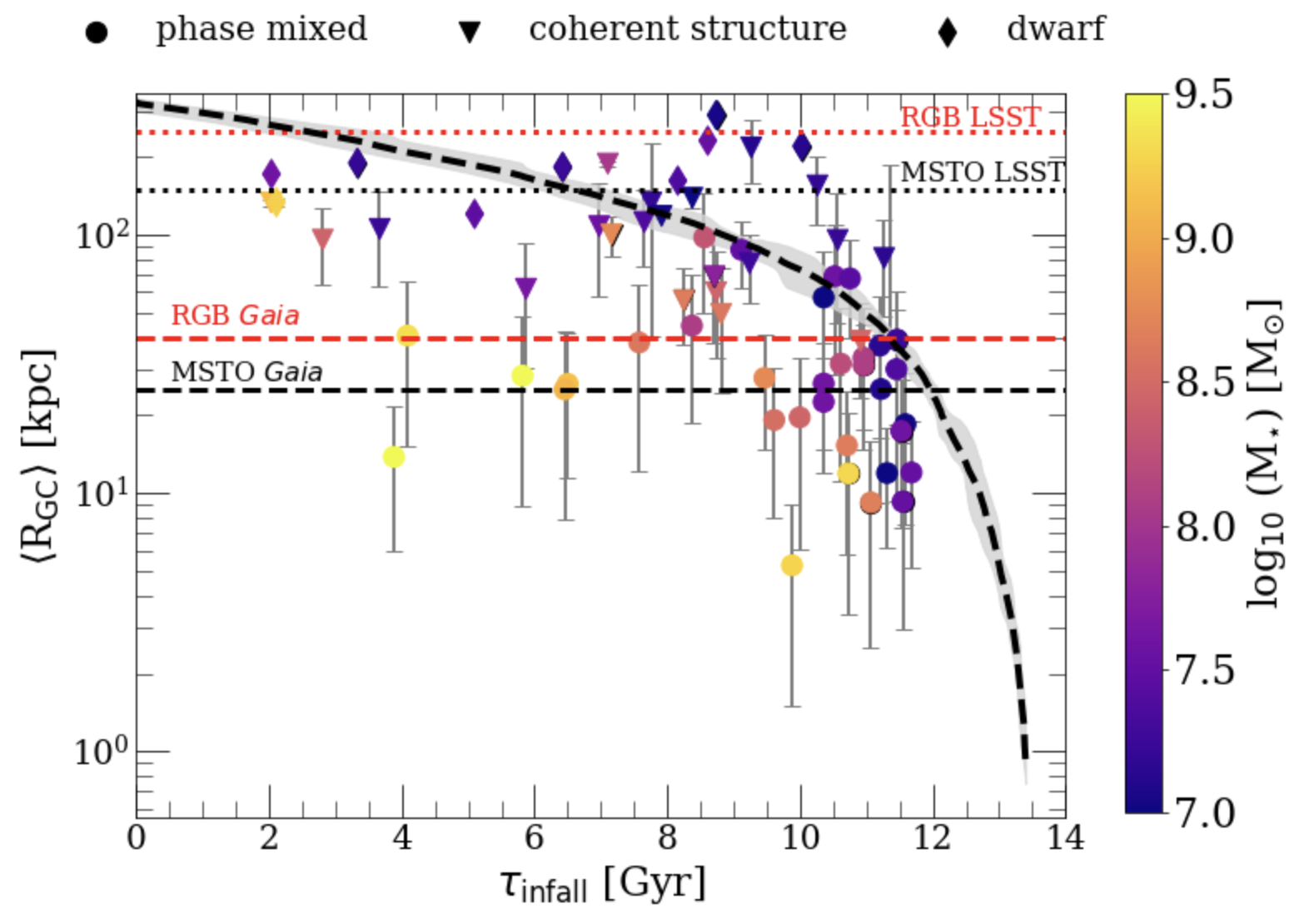}
\caption{Mean Galactocentric radius value (at $z$ = 0) for every accretion event as a function of their merger infall time (where $\tau_{\mathrm{infall}}$=0 corresponds to present day), color coded by the stellar mass of each subhalo at the time of infall. The error bars indicate the mean of the 16$^{th}$ and 84$^{th}$ percentile ranges of the star particles. The dashed black line illustrates the mean value of the evolution of the virial radius for the seven Milky Way-like halos used, with the shaded region demarking the 16$^{th}$ and 84$^{th}$ percentile range. Overplotted are predictions for observing detection limits for individual stars within the Milky Way Galaxy, from upcoming massive surveys from \citet[][]{Sanderson2019}, assuming the faintest ($r$$\sim$20.5-24.5) limiting magnitudes for RGB and MSTO stars. At large $\tau_{\mathrm{infall}}$, galaxies of all stellar masses deposit the bulk of their stars close to the Galactic centre of the host (MW-like) Galaxy. However, at fixed $\tau_{\mathrm{infall}}$, more massive systems tend to deposit the bulk of their stars closer to the host Galaxy's centre. Furthermore, our results suggest that the smaller mass systems (M$_{\star}$ $<$ 10$^{8}$ M$_{\odot}$) that are accreted at intermediate to low $\tau_{\mathrm{infall}}$ can be contained at large Galactocentric distances, of over 150 kpc, that will only be detectable with upcoming surveys like LSST.}
    \label{rgc_tinfall}
\end{figure*}

Fig~\ref{rgc_tinfall} shows the mean Galactocentric radius ($\langle$R$_{\mathrm{GC}}\rangle$) value at $z$=0 of star particles belonging to every accretion event identified in this work as a function of $\tau_{\mathrm{infall}}$, and color coded by their stellar mass. The error bars shown in this plot show the average of the difference between the 16$^{th}$ and 84$^{th}$ percentile ranges for the full distribution of each object. The black line defines the mean value for the evolution of the virial radius for the seven Milky Way-like hosts as a function of lookback time.

Our results reveal a clear relation between $\langle$R$_{\mathrm{GC}}\rangle$, stellar mass, and $\tau_{\mathrm{infall}}$. Here, lower (stellar) mass mergers deposit the bulk of their stars at higher $\langle$R$_{\mathrm{GC}}\rangle$ than more massive mergers at fixed $\tau_{\mathrm{infall}}$. We find this to also be the case for total mass (defined in Table~\ref{definitions} as M$_{\mathrm{DM}}$). For the most massive mergers (M$_{\star}>$ 10$^{8}$M$_{\odot}$), the floor $\langle$R$_{\mathrm{GC}}\rangle$ value increases with decreasing $\tau_{\mathrm{infall}}$ as the host MW-like halo grows, growing from a value of $\langle$R$_{\mathrm{GC}}\rangle<$ 10 kpc at $\tau_{\mathrm{infall}}>$ 10 Gyr to a value of $\langle$R$_{\mathrm{GC}}\rangle >$ 30 kpc at $\tau_{\mathrm{infall}}<$ 4 Gyr. Along similar lines, we find that accretion events that deposit the bulk of their stars at higher Galactocentric values ($\langle$R$_{\mathrm{GC}}\rangle>$ 40 kpc) are predominantly low mass (M$_{\star}<$ 10$^{7.5}$M$_{\odot}$), unless they get accreted closer to $z$=0 (i.e., low $\tau_{\mathrm{infall}}$), and are all either coherent streams or bound dwarfs mergers. These accretion events have long dynamical times, allowing the debris from these mergers to remain kinematically coherent in phase-space. Conversely, mergers that deposit the bulk of their stars closer to the Galactic centre of the MW-like host are predominantly phase-mixed. 

It is particularly interesting to ask where we might find the remnants of low mass, long-dead satellites that would be below the detection limits of higher redshift studies. At high infall times (namely, $\tau_{\mathrm{infall}}>$ 11 Gyr) we find a large number of lower mass accretion events (M$_{\star}<$ 10$^{7.5}$M$_{\odot}$) that range in dynamical state and are present across wide range of mean Galactocentric distances. Consistent with prior work, our results demonstrate that the inner regions of the stellar halo should host phase mixed debris from these objects. In addition, a significant fraction of our low-mass sample are present at high $\langle$R$_{\mathrm{GC}}\rangle$ values. Moreover, these remain coherent in phase space. While errors in distances and decreasing surface density will impede the detectability of these substructures spatially at large R$_{\mathrm{GC}}$, these kinematically cold features will be promising to detect in kinematic surveys. Therefore, these simulations suggest that we are not limited to local volumes to find remnants of very early and low mass accretion events, and point to a great deal of discovery potential for these systems with future astrometric and spectroscopic surveys.  

Along those lines, the error bars in Fig~\ref{rgc_tinfall} (which bracket the 16$^{th}$--84$^{th}$ range) provide information of the distribution across radii for each disrupted satellite. Given our findings, mergers that deposit the bulk of their stars at the closest Galactocentric distances (namely, $\langle$R$_{\mathrm{GC}}\rangle$ $\lesssim$ 30 kpc) have a relatively small, yet noticeable, spread in the Galactocentric distances at which they deposit these stars, ranging from anywhere between a range of 5 kpc to 30 kpc. Conversely, the majority of accretion events that end up depositing the bulk of their stars at large Galactocentric distances (namely, at $\langle$R$_{\mathrm{GC}}\rangle$ $>$ 100 kpc) can be divided into two subcategories. The first is comprised by primarily coherent stream accretions, which have very high spread values (namely, a range greater than $\sim$30 kpc). The second group is populated primarily by accretion events classified as present-day dwarf galaxies, that present very low range, on the order of $\sim$10 kpc. 

We note that there are four coherent stream mergers that present lower spread values, on the order of $<$ 15 kpc. We examined the distribution of these disrupted satellites in the X-Y and X-Z projections, and found that these mergers present lower spreads in their Galactocentric radii distribution because they present thin and long tidal tails and/or have a significant amount of mass still bound in their progenitor system that has not fully disrupted. This spatial distribution would lead to these debris occupying large Galactocentric distance values, whilst still maintaining a low spread. Conversely, there is one coherent stream merger that presents a very large Galactocentric radii distribution, on the order of $\sim$90 kpc. When examining the distribution of star particles belonging to this disrupted satellite in the X-Y and X-Z planes, we found that this accretion event presents a very radial and extended distribution, thus leading to the large Galactrocentric radius spread value determined for its given average Galactocentric distance. 

In order to contextualize our findings with respect to current/upcoming surveys, we overplot in Fig~\ref{rgc_tinfall} observing limits estimated from \citet{Sanderson2019} for different stellar tracer populations in the Milky Way. Specifically, we show as red/black lines the assumed limiting magnitude limits for red giant branch (RGB)/main sequence turn off (MSTO) stars, two stellar tracer populations that will be advantageous to trace due to their bright luminosity and populous numbers, respectively (\citealp[e.g.,][]{Cunningham2019a,Cunningham2019b}). Dashed lines indicate the distance limit for samples of RGB (shown in red) and MSTO (in black) stars observed by \textit{Gaia} mission \citep[][]{Gaia2018} and with 4-m class telescope spectroscopy. Dotted lines show estimates of the distance limits for these tracers observed with Rubin Observatory's Legacy Survey of Space Time \citep[LSST;][]{Ivezic2019}. In Section~\ref{implications} we discuss in detail the implications of our findings in the context of observational work. However, in brief, our results suggest that current astrometric and spectroscopic surveys are only scratching the surface of the debris from cannibalised satellite galaxies that are predicted to inhabit the stellar halo of the Galaxy, and are likely biased to only identifying the most massive and/or the most recently accreted disrupted galaxies. This reasoning is two-fold: 1) more massive systems will dominate the local (observable) stellar halo, and thus will likely be more easily detectable based on sheer numbers; 2) more recently accreted disrupted satellites (e.g., Sagittarius dSph) will still retain easily detectable coherent stream debris, making them also more easily detectable.
The advent of large-scale stellar surveys like LSST, WEAVE, and 4MOST, with its ability to observe stellar tracer populations with high numbers at far larger Galactocentric distances than current surveys, will help make possible such discoveries. 

\section{Orbital properties}
\label{kinematics}
Having studied the resulting spatial distribution of the identified accretion events, in this Section we discuss the distribution of their orbital properties. Specifically, we set out to interpret the resulting mean orbital energy, angular momentum, and circularity values (as defined in Table~\ref{definitions}) for every merger event as a function of both infall time and stellar mass. As integrals of motion (such as orbital energy and actions) are, under the assumption of a static potential and potential symmetry, invariants, the results obtained in this section will provide insight on how different types of accretion events manifest themselves in key diagnostic orbital planes that can be qualitatively compared to large scale observations of stars in the MW (see for example \citet[][]{Santistevan2022}).

\subsection{Orbital energy}
\label{sec_energy}
\begin{figure*}
\centering
\includegraphics[width=\textwidth{}{}]{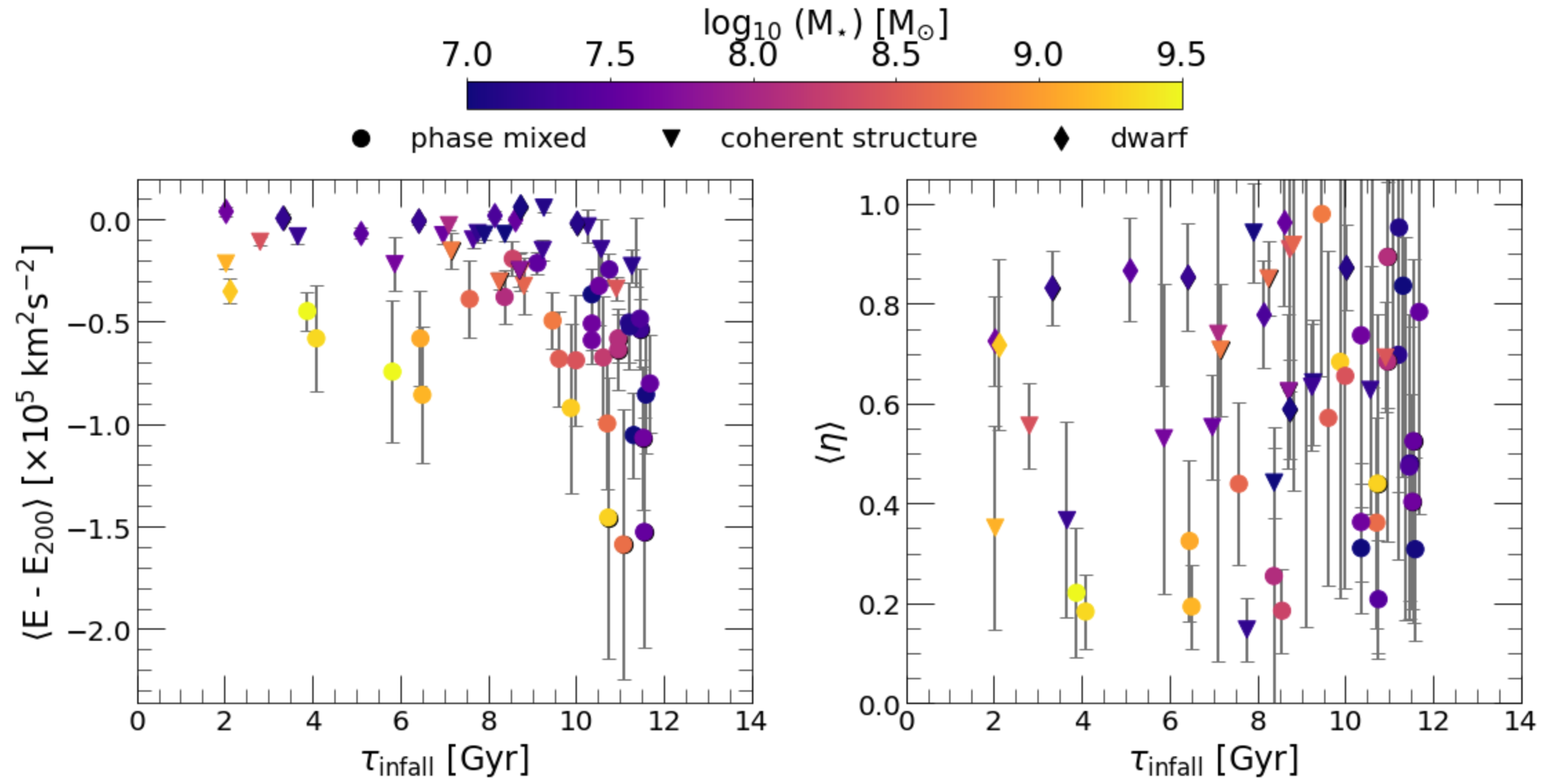}
\caption{Normalized mean orbital energy (left) and mean orbital circularity (right) for every accretion event (w.r.t. their host Galaxy's potential) at $z$=0 as a function of $\tau_{\mathrm{infall}}$, color coded by the respective stellar mass. As in Fig~\ref{rgc_tinfall}, the errorbars illustrate the 16$^{th}$ and 84$^{th}$ percentile range. }
    \label{orbits_tinfall}
\end{figure*}

We begin by examining the resulting (normalized) mean orbital energy in the left panel of Fig~\ref{orbits_tinfall} as a function of $\tau_{\mathrm{infall}}$, where each merger event is color coded by its respective stellar mass. Here, the mean orbital energy (E) is the mean
kinetic and potential energy per unit mass over all particles that belong to each accretion event at $z$=0 (namely, E = $\Phi$ + $K$, where $\Phi$ and $K$ are the mean potential energy per unit mass and mean kinetic energy per unit mass, respectively), normalized to the mean orbital energy value at $R$=200 kpc (which we denoted as E$_{200}$). This yields the mean normalized orbital energy $\langle$E $-$ E$_{200}\rangle$. Furthermore, the potential ($\Phi$), taken from \citet[][]{Arora2022}, is comprised of three components: a NFW profile for the dark matter halo \citep[][]{Navarro1996}, a spherical potential model for the central bulge, and a Miyamoto-Nagai profile modeling the cylindrical disk of the simulation \citep[][]{Miyamoto1975}. This potential is then propagated in simulation time as a time-evolving multipole potential, modeled using a low-order multipole expansion assuming axisymmetry, evaluated at each snapshot independently. For the case of the kinetic energy term, we determine this quantity per unit mass as $K$ = $\frac{1}{2}v^{2}$.

Figure~\ref{orbits_tinfall} shows that mean orbital energy is strongly dependent on the stellar mass and infall time of a disrupted satellite, and follows a similar relation to that observed in Fig~\ref{rgc_tinfall}. This result is not surprising, given that the potential energy is inversely proportional to galactocentric radius. We find that at early times ($\tau_{\mathrm{infall}}$ $\gtrsim$ 11 Gyr) accretion events have a wide range of orbital energies, spanning from --1.75 $\times10^{5}$ $<$ $\langle$E -- E$_{200}\rangle$ $<$ 0.05$\times10^{5}$ km$^{2}$s$^{-2}$, whereas at more recent times (namely, $\tau_{\mathrm{infall}}$ $\lesssim$ 7 Gyr) accretion events have a smaller range in mean orbital energies (--0.75 $\times10^{5}$ $<$ $\langle$E -- E$_{200}\rangle$ $<$ 0.05$\times10^{5}$ km$^{2}$s$^{-2}$). At fixed infall time, higher-mass accretion events deposit the bulk of their stars at lower orbital energies than their lower-mass counterparts, with the minimum orbital energy increasing over time due to the growth of the MW-like host halo potential. Furthermore, we find that for fixed $\tau_{\mathrm{infall}}$, accretion events that deposit the bulk of their stars at lower energies are primarily classified as phase-mixed mergers. Conversely, we find the mergers at the highest orbital energies ($\langle$E -- E$_{200}\rangle$ $\gtrsim$ --0.15$\times10^{5}$ km$^{2}$s$^{-2}$) to be primarily either coherent stream and/or present-day dwarf galaxies.

The spread in the orbital energy values, defined as the range between the 16$^{th}$ and 84$^{th}$ percentiles, are also shown in the left panel of Fig~\ref{orbits_tinfall} as error bars. Here, we see that accretion events at earlier $\tau_{\mathrm{infall}}$ have a much wider range in the spread of their orbital energy values. At fixed infall time, we find that more massive accretion events have on average higher spreads in their energies than their lower mass counterparts. Moreover, we also see that the value of the spread decreases with decreasing $\tau_{\mathrm{infall}}$. This is likely due to the combination of two effects: firstly, the potential of the host changes over time, and these changes are more drastic at early times, causing the orbital energies of the stars of the disrupted satellite to occupy a wider range of values; secondly, the value of the potential energy at R$=200$ kpc increases over time due to the overall growth of the host halo, and thus the dynamical times are longer. This results in later accreted mergers taking longer to phase mix, and having less time to do so. Both the change in the potential of the host, as well as the mass ratio between the host and the disrupted galaxy, are mechanisms that result in debris from recent accretion events having lower spreads in their orbital energies.

\subsection{Circularity}

We now focus on the mean circularity ($\langle \eta \rangle$)\footnote{These values were computed using the potentials determined in \citet{Arora2022}, utilizing the \texttt{AGAMA} \citet[][]{Vasiliev2019_agama} galactic dynamics package.} values for star particles associated with all the accretion events studied in this work. As described in Table~\ref{definitions}, $\eta$ is defined as the ratio of the total angular momentum of a merger with the angular momentum of a circular orbit with the same orbital energy. More explicitly, the circularity term was calculated as a weighted sum of the angular momenta (and respective L$_{\mathrm{circ}}$ values) for each star particle in each merger, and can be defined analytically as:
\begin{equation}\label{eq2}
    \langle\eta\rangle = \Bigg \langle \frac{\mathrm{L_{tot}(E)}}{\mathrm{L_{circ}(E)}} \Bigg \rangle,
\end{equation}
where L$_{\mathrm{tot}}$(E) and E are the angular momentum per unit
mass and orbital energy per unit mass, respectively, for every star particle associated with an accretion event, and L$_{\mathrm{circ}}$(E) is the angular momentum per unit mass of a circular orbit with the same E (namely, L$_{\mathrm{circ}}$(E) = $r_{\mathrm{circ}}$(E) $\times$ $v_{\mathrm{circ}}$(E), see Table~\ref{definitions} for further details). Thus, $\langle \eta \rangle$ = 1 corresponds to a perfectly circular orbit at $z$=0, whilst $\langle \eta \rangle$ = 0 corresponds to a perfectly radial orbit. 
We note that there are two mergers, one in m12b and one in m12m, for which we obtained a circularity value of $\langle \eta \rangle$ $>$ 1 (1.28 and 3.12, respectively) which is physically unreasonable. \citet[][]{Panithanpaisal2021} argue that the time-dependent and non-spherically symmetric potential, especially at early times, is the culprit for such problematic circularity values. Thus, we choose to exclude these mergers when examining the resulting circularities of accretion events in the set of MW-like galaxies studied. As we only remove two mergers from our sample, we are confident that the exclusion of these accretion events will not impact any conclusions derived from our results.

The right panel of Fig~\ref{orbits_tinfall} displays the resulting mean orbital circularity values obtained for each accretion event as a function of $\tau_{\mathrm{infall}}$, color coded by stellar mass. The error bars in this plot show the spread in the orbital energy and circularity values, defined as the 16$^{th}$--84$^{th}$ percentile ranges for each object. From our results, we find that there is no clear relation between $\langle \eta \rangle$ with either infall time nor stellar mass, as mergers of different $\tau_{\mathrm{infall}}$ and M$_{\star}$ present a wide variety of $\langle \eta \rangle$ values. Interestingly, we find that on average, phase mixed accretion events present lower $\langle \eta \rangle$ values when compared to those mergers classified as coherent streams and/or dwarfs. This result is in agreement with \citet[][]{Johnston2008}, and suggests that surviving satellites are on more circular orbits than their phase mixed counterparts (see also \citet[][]{Santistevan2022}). However, we do note that there is a spread, and that there are a small number of phase mixed mergers that present higher circularity values and a small number coherent stream/dwarfs mergers that present circularity values on the order of $\langle \eta \rangle$$\sim$0.2.

\subsection{The link between orbital energy and circularity}
\label{sec_elz}

\begin{figure*}
\centering
\includegraphics[width=\textwidth]{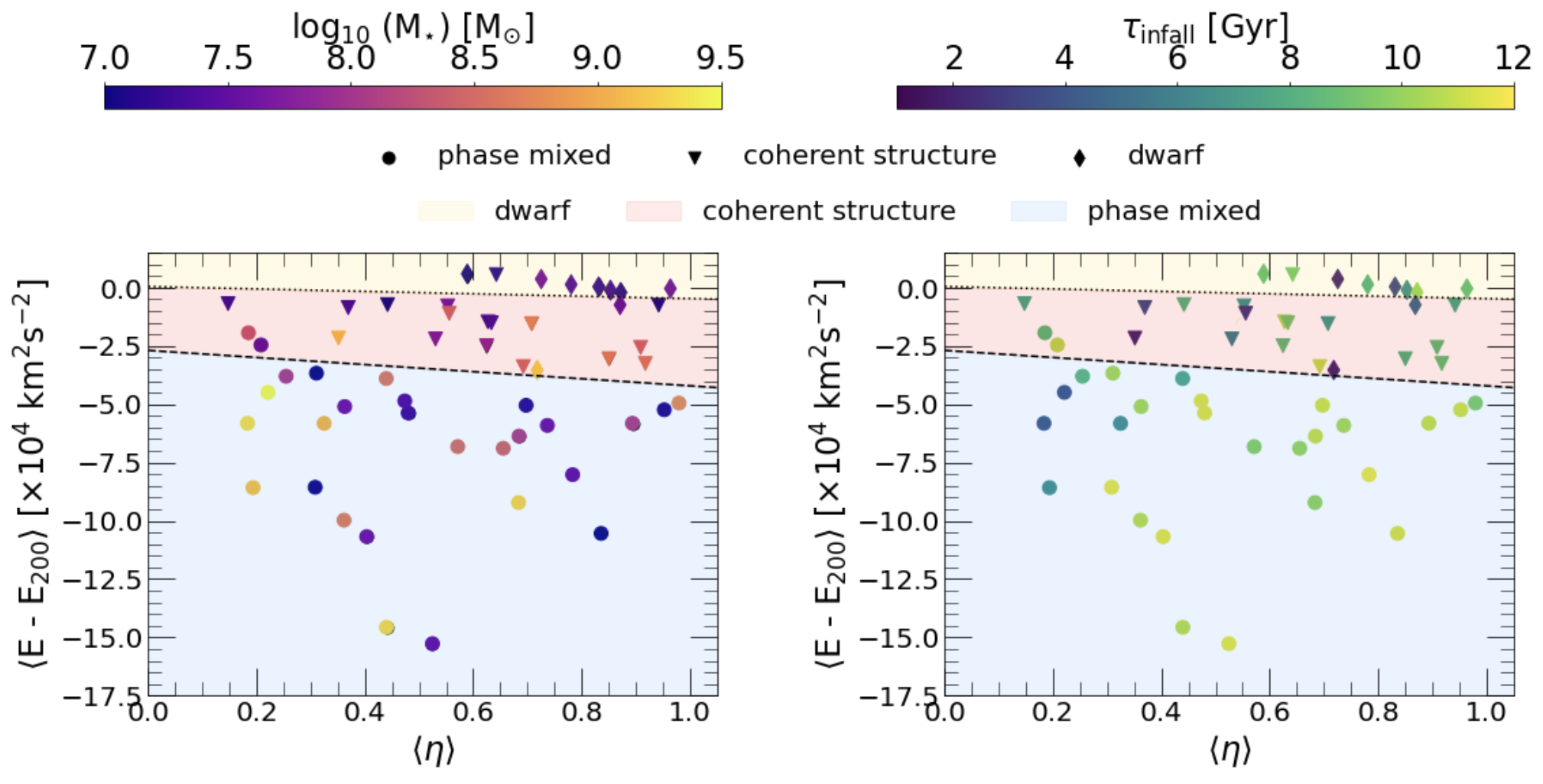}
\caption{Mean orbital energy as a function of mean circularity for all accretion events studied in this work, color coded by stellar mass (left) and infall time (right). Indicated in these plots are three groups (shaded regions) corresponding to the predictions determined using the support vector machine classifier \texttt{sklearn.svm.SVC} (see text for details). The slope and intercept values of the two division lines are slope:--1.52(--0.52); intercept:--2.69(0.05) for the dashed(dotted) lines, respectively. It is possible to use the mean orbital energy and circularity values of a disrupted galaxy to statistically predict its present day dynamical state.}
    \label{energy_circularity}
\end{figure*}

Having examined all the orbital properties of accretion events in MW-like halos independently, we now set out to study the position of such mergers in a combination of orbital parameters. We show the resulting mean orbital energy values for each accretion event as a function of their mean circularity in Fig~\ref{energy_circularity}. In addition, we quantify the distinction of different dynamical state classes of accretion events by training a linear kernel Support Vector Machine (SVM) classifier using the \texttt{sklearn.svm.SVC} routine and utilising as input the orbital energy and circularity values of every merger. Specifically, we set the kernel to be linear and define the regularization parameter to 1. Our training set comprises of all mergers in the \texttt{m12b}, \texttt{m12c}, \texttt{m12f}, \texttt{m12i}, and \texttt{m12m} simulations, and our validation set is comprised of all mergers in the \texttt{m12r} and \texttt{m12w} simulations. We perform this test for two independent samples: firstly, for phase mixed and coherent structure mergers; and secondly, for coherent structure and dwarf disrupted satellites. The results from this SVM classifier yield two linear trends able to distinguish accretion events in Fig~\ref{energy_circularity} based on dynamical state, and are shown as the shaded regions in this diagram.

From our results, it is evident that an accretion event's present-day dynamical state is clearly dependent on its orbital energy and circularity, as the dwarf, coherent stream, and phase mixed mergers occupy different loci in this plane. In detail, we find that the dwarf mergers occupy a position at high orbital energies and high circularities, congregating in a narrow locus at the top-right corner of the diagram. Conversely, we find that phase mixed mergers, which present a larger range in mean orbital energies, present a wide range of mean circularity values. However, these types of accretion events all sit at lower orbital energies when compared to their dwarf counterparts. Lastly, occupying a locus in between the dwarf and phase mixed mergers, we find the coherent streams, which present relatively high orbital energies and a wide range of circularities. This projection (or view) of orbital parameters usefully summarizes the dependence of stellar mass and infall time on the resulting mean orbital energy and circularity (and therefore by definition, also with angular momentum). It also reveals the clear dependence of a merger's dynamical state with these pivotal parameters, and illustrates that one does not require an infall time or stellar mass to determine a mergers morphology at present day (as infall time and stellar mass are correlated with orbital energy). Of particular importance is the fact that this orbital plane illustrates the wide variety of orbits mergers of similar mass and infall time can adopt, and how dependent these distributions are on the initial conditions of the accretion event.

\section{Chemical properties}
\label{Chemistry}

We next examine the position of the mergers in the $\alpha$-Fe plane. As the chemistry of a system should remain constant over time after a satellite is consumed by a larger host given that its star formation ceases due to gas being stripped (\citealp[e.g.,][]{Robertson2005,Font2006}), the results from this examination should serve as a strong diagnostic to distinguish different types of accretion events.  
 
The distribution of stars in the [$\alpha$/Fe]--[Fe/H] is an indicator of the chemical evolution and star formation history (SFH) of the population. Understanding the [$\alpha$/Fe] and [Fe/H] abundances of stars associated with a galaxy can help place strong constraints on its formation and evolution. When applied to stars from satellites accreted by a MW-like galaxy, their distribution in this chemical plane is a clear diagnostic of the time in which the accreted satellite ceased star formation (\citealp[e.g.,][]{Robertson2005,Font2006,Mackereth2019, Khoperskov2022}). It also provides an opportunity to study how lower-mass galaxies chemically evolved in the early Universe \citep[e.g.,][]{Corlies2018}. For example, \citet[][]{Panithanpaisal2021} and \citet[][]{Cunningham2022} showed how the progenitors of low to intermediate-mass streams differed chemically from the present-day population of satellite galaxies. This result emphasizes the ability to use these simulations to gain a deeper understanding of how chemical evolution in low- and intermediate mass galaxies has varied across cosmic time.

\begin{figure*}
\centering
\includegraphics[width=\textwidth{}{}]{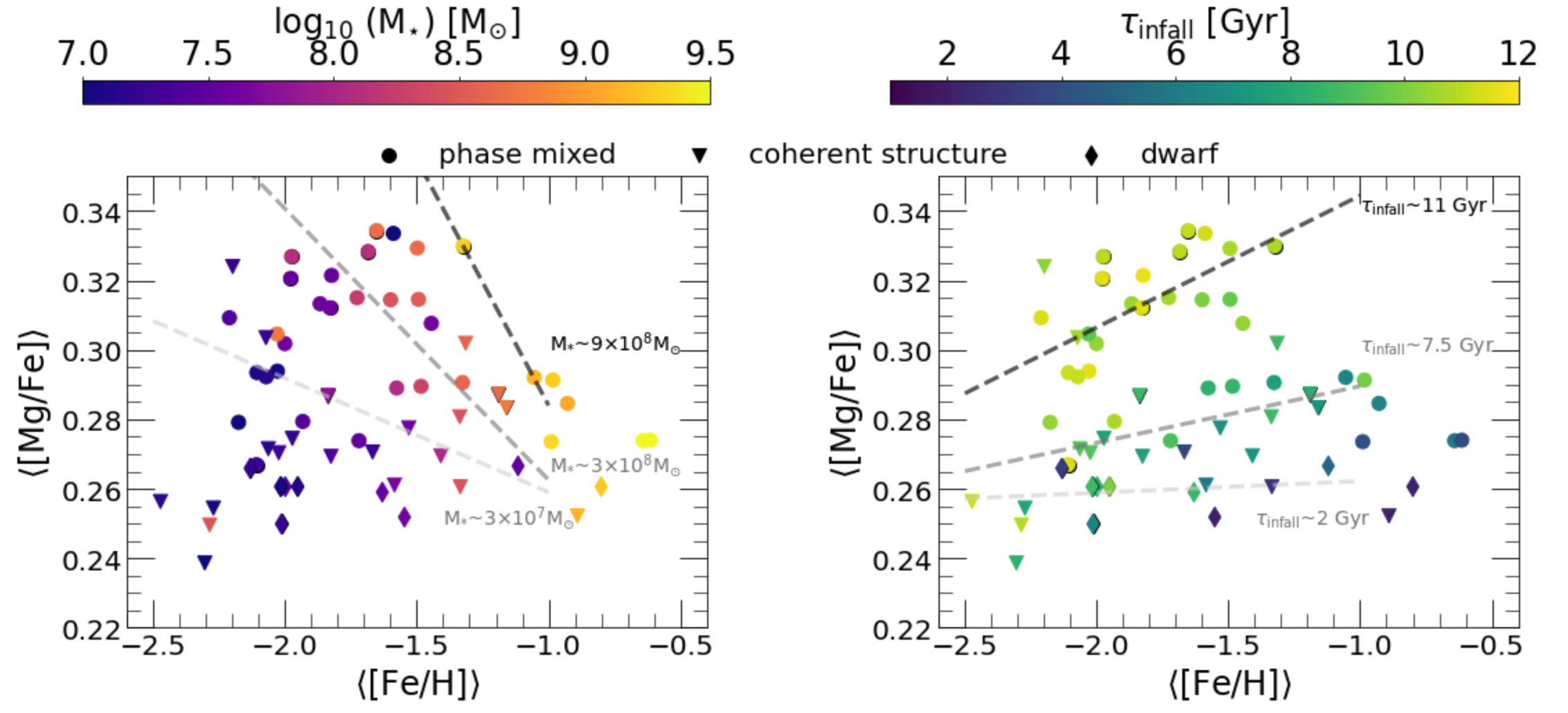}
\caption{Mean [Mg/Fe] and [Fe/H] values for all the accretion events identified in this work, color coded by stellar mass (left) and infall time (right). It is evident that these chemical properties are intimately related with the progenitors stellar mass and infall time. To illustrate this point, we show the iso-mass (namely, for M$_{\star}$=3$\times$10$^{7}$/3$\times$10$^{8}$/9$\times$10$^{8}$ M$_{\odot}$) and iso-age (namely, for $\tau_{\mathrm{infall}}$=2/7.5/10 Gyr) trends in (light-grey, grey, black). For example, more massive and/or earlier mergers on average present higher mean [Mg/Fe] values at fixed [Fe/H]. The average spread values, defined as the mean of the 16$^{th}$ and 84$^{th}$ percentiles, is [Mg/Fe]$\pm$0.06 and [Fe/H]$\pm$0.33 for phase mixed debris, [Mg/Fe]$\pm$0.08 and [Fe/H]$\pm$0.39 for coherent structure debris, and [Mg/Fe]$\pm$0.08 and [Fe/H]$\pm$0.40 for dwarf/satellites.}
    \label{mgfe}
\end{figure*}

In Figure \ref{mgfe}, we show the mean iron metallicity versus the mean magnesium abundance ([Mg/Fe]; we use Mg as our $\alpha$ element) color coded by their stellar mass (left), and by their infall time (right). We note that the average spread values, defined as the mean of the difference between the 84$^{th}$ and 16$^{th}$ percentiles, is [Mg/Fe]$\pm$0.06 and [Fe/H]$\pm$0.33 for phase mixed debris, [Mg/Fe]$\pm$0.08 and [Fe/H]$\pm$0.39 for coherent structure debris, and [Mg/Fe]$\pm$0.08 and [Fe/H]$\pm$0.40 for dwarf/satellites.

Our results show that there is a strong dependence of the position of the bulk of the stars of accretion events in the [Mg/Fe]--[Fe/H] plane with stellar mass and infall time, that subsequently leads to a relationship with its dynamical state. Focusing on the left panel of Fig~\ref{mgfe}, we find that more massive accretion events have a higher mean [Fe/H] for fixed mean [Mg/Fe] when compared to their lower mass counterparts, and that such relation is consistent across all $\langle$[Mg/Fe]$\rangle$ values. This effect is due to the mass-[Fe/H] relationship of galaxies \citep[][]{Kirby2013,Hidalgo2017}, where those satellite galaxies that are more massive will have enriched their ISM with more metals when compared to the lower-mass counterparts. Similarly, at constant $\langle$[Fe/H]$\rangle$, we find that lower mass accretion events present overall lower mean [Mg/Fe] when compared to their higher stellar mass siblings (see also \citet[][]{Samuel2022}). This difference is due to the lower-mass systems overall having lower and/or specific star formation rates, which subsequently leads to less SNII explosions per unit mass and a lower proportional production of $\alpha$ elements and Fe (see Mason et al, in prep). If one now focuses on the right panel of Fig~\ref{mgfe}, where each merger is color coded by their respective $\tau_{\mathrm{infall}}$ value, our results reveal another set of trends. Here we find that, for fixed mean [Fe/H], disrupted satellites that get engulfed by their host MW galaxies at earlier times (i.e., high $\tau_{\mathrm{infall}}$) display higher mean [Mg/Fe] values when compared to accretion events that get accreted at later times (i.e., lower $\tau_{\mathrm{infall}}$ values). Additionally, we find that those accretion events classified as phase mixed typically present higher $\langle$[Mg/Fe]$\rangle$ values for fixed $\langle$[Fe/H]$\rangle$ when compared to those mergers classified as coherent streams and/or dwarfs. We reason that this is due to phase mixed accretion events generally being quenched earlier when compared to coherent structures and/or dwarf satellites. Therefore, for phase mixed debris to reach a given stellar mass (and therefore a given [Fe/H]), their star formation rate must be higher/more efficient than in lower stellar mass at similar [Fe/H]. Conversely, the latter will have a lower, but more prolonged, star formation history (this was also shown in detail in a recent work by \citet[][]{Cunningham2022} using the same suite of simulations, Fig 3).

We note that the relationship between $\langle$[Mg/Fe]$\rangle$, $\langle$[Fe/H]$\rangle$, and peak stellar mass for the accretion events identified in this work reveal diagonal iso-mass trends in this plane, where for fixed stellar mass, an accretion event can be positioned in the [Mg/Fe]-[Fe/H] plane along a diagonal line that decreases in [Fe/H] with increasing [Mg/Fe]. We also note that such iso-mass trends observed in the left hand side of Fig~\ref{mgfe} are dependent on $\tau_{\mathrm{infall}}$, where those accretion events of a given stellar mass that are accreted at higher $\tau_{\mathrm{infall}}$ values present a higher mean [Mg/Fe] and lower mean [Fe/H] when compared to mergers of similar mass and smaller $\tau_{\mathrm{infall}}$. 

The intricate relationships found between these key parameters are intimately linked to the star formation history and chemical evolution of the accreted satellite galaxy. Our results have confirmed previous expectations (\citealp[][]{Robertson2005,Font2006,Johnston2008,Mackereth2019}) that for fixed mean [Fe/H], more massive and/or earlier accreted systems present a higher mean [Mg/Fe] value when compared to their lower mass and/or later accreted siblings. At fixed $\tau_{\mathrm{infall}}$, this is likely because more massive systems have had a larger number of supernovae (SN) contributions, which pollute the ISM with (primarily) $\alpha$ elements (for the case of core-collapse SN --SNII--) and Fe (for the case of SN type Ia --SNIa--), due to their higher star formation rate (SFR) when compared to smaller mass systems, which have a slower SFR and smaller number of SN explosions.

Along those lines, for fixed stellar mass, those systems that merged at an earlier lookback time (namely, high $\tau_{\mathrm{infall}}$ values) would have had their star formation quenched earlier. This results in these earlier mergers having a shorter time to evolve chemically, yielding a resulting position in the [Mg/Fe]-[Fe/H] plane of higher mean [Mg/Fe] and lower mean [Fe/H] values when compared to their later accreted and similar mass counterparts, that would have had more time to evolve chemically and enrich their ISM with more heavier metals (like Fe) from delayed SNIa, diluting the [Mg/Fe] ratio whilst increasing [Fe/H].

The results obtained from examining the mean [Mg/Fe] and [Fe/H] abundances disclose a relationship between an accreted population's chemistry and their corresponding stellar mass and infall time. This result is not surprising, given that the chemical evolution of a galaxy will be governed by the amount of time such system is able to evolve before being accreted (and ceasing star formation), and by the amount of gas such system is able to turn into stars (i.e., its star formation), which in turn is intimately linked to the mass of the system. In more detail, the iso-mass and iso-age relations observed for the accretion events studied in Fig~\ref{mgfe} are a clear manifestation of the evolution of the mass-metallicity relationship of galaxies across time. To summarise, the clear relationship revealed between the $\langle$[Mg/Fe]$\rangle$ and $\langle$[Fe/H]$\rangle$ abundances, the stellar mass, and the infall time of each accretion event is strong, and serves as a powerful tool for disentangling the properties of such mergers.

As the metallicity in the FIRE simulations does not precisely match the observations \citep[][]{Escala2018, Wheeler2019}, we note that these results should be taken as qualitative. Our findings on the chemical properties of these disrupted satellites should not be used as a quantitative measure in order to predict the stellar mass and infall time of debris identified in the stellar halo of the MW conjectured to originate from cannibalised satellites. Instead, we suggest that the iso-mass and iso-age relations shown in Fig~\ref{mgfe} should be used as qualitative blueprint for understanding how different accretion events manifest themselves in a fundamental chemical composition plane, as the general trends and relationships with the chemical properties of these systems are still viable. 

Furthermore, since particle initialisation in the FIRE-2 models prescribes star particles to be born with a primordial [Fe/H]$=-4$ and [$\alpha$/Fe]$=$0, disrupted satellites that present a lower $\langle$[Fe/H]$\rangle$ value will on average display a much lower $\langle$[Mg/Fe]$\rangle$ than expected (see for example \citet[][]{Font2006}). This is what causes the decline in the trend of $\langle$[Mg/Fe]$\rangle$ below $\langle$[Fe/H]$\rangle$ $<$ --2 in Fig~\ref{mgfe}, and causes these disrupted satellites to present un-physically low $\langle$[Mg/Fe]$\rangle$ values. While this model prescription serves as an added nuisance, we argue that it does not impact the iso-mass and iso-age trends towards higher $\langle$[Fe/H]$\rangle$.

\section{Summary and Conclusions}
\label{conclusion}

In this paper we have examined accretion events in a set of seven MW-like halos using the \textit{Latte} suite of FIRE-2 zoom-in cosmological simulations with the goal of characterizing the relationships between a merger's properties (stellar mass and infall time) and those of the stellar debris it leaves behind.  We have explored a range of diagnostic and observable planes, including their spatial distribution (Section~\ref{spatial_distribution}), orbital properties (Section~\ref{kinematics}), and chemical properties (Section~\ref{Chemistry}). 
A summary of the number of disrupted satellites and their properties is listed in Table~\ref{merger_summary}.
Our analyses aimed to address three questions: "\textit{What are the properties of disrupted satellites}?", "\textit{Do our expectations of how satellite galaxies get disrupted from tailored/idealised $N$-body simulations survive in a cosmological context}?" and "\textit{What are the implications for observations}".

\subsection{What are the properties of disrupted satellites?}

\subsubsection{Spatial distribution}

\begin{itemize}
\item The spatial distribution of stars deposited by disrupted satellites disclose a clear relationship between stellar mass, infall time, and $\langle$R$_{\mathrm{GC}}$$\rangle$, where more massive mergers deposit the bulk of their stars at lower mean Galactocentric distances when compared to lower mass mergers at fixed $\tau_{\mathrm{infall}}$. Moreover, we find that the minimum $\langle$R$_{\mathrm{GC}}$$\rangle$ value increases with decreasing $\tau_{\mathrm{infall}}$ as the host MW-like halo grows, growing from $\langle$R$_{\mathrm{GC}}$$\rangle$ $<$ 10kpc at $\tau_{\mathrm{infall}}$ $>$ 10Gyr to $\langle$R$_{\mathrm{GC}}$$\rangle$ $>$ 30kpc at $\tau_{\mathrm{infall}}$ $<$ 4Gyr (for the most massive mergers M$_{\star}$ $>$ 10$^{9}$M$_{\odot}$). 

\item Disrupted satellites that deposit the bulk of their stars at higher mean Galactocentric distances (namely, $\langle$R$_{\mathrm{GC}}$$\rangle$ $>$ 40kpc) present a stellar mass below M$_{\star}$ $<$ 10$^{7.5}$ M$_{\odot}$, and are classified primarily as either coherent stream/ bound dwarf mergers. This result suggests that there is likely a great deal of undiscovered structures/satellites in regions of the Galactic stellar halo beyond $\sim$30 kpc, that should be identifiable with future large-scale stellar surveys (e.g., \textit{Gaia}, DESI, WEAVE, 4MOST). This will be particularly exciting for the Rubin/LSST survey, which will be able to detect streams/dwarf galaxies to distances beyond R$_{\mathrm{GC}}$ $>$ 50 kpc \citep[][]{Sanderson2019}. 

\item Conversely, accretion events that present lower mean Galactocentric distances (namely, $\langle$R$_{\mathrm{GC}}$$\rangle$ $<$ 40kpc) are primarily massive (namely, M$_{\star}$ $>$ 10$^{8.5}$ M$_{\odot}$) phase mixed mergers. 

\subsubsection{Orbital properties}

\item The resulting position occupied by different accretion events in the orbital energy plane studied is strongly dictated by the intimate relationship between a merger's peak stellar mass and infall time. The interplay between these two fundamental parameters governs a disrupted satellite's orbital energy at present day, and plays a strong role on the dynamical state a merger will adopt at $z$=0. The left panel of Fig~\ref{orbits_tinfall} shows that for fixed $\tau_{\mathrm{infall}}$, more massive accretion events present overall lower orbital energy values when compared to their lower mass counterparts. This relation is consistent across all infall times, but however the magnitude of the minimum mean orbital energy of a merger increases with decreasing $\tau_{\mathrm{infall}}$, due to the growth of the host MW-like halo over time. 

\item The dynamical state of a disrupted satellite is also influenced by the system's stellar mass and infall time, since most of the more massive accretion events are typically phase mixed mergers, whereas lower mass accretion events are primarily bound coherent streams/dwarfs.

\item Mergers with lower mean orbital energy values (namely, more massive and typically phase mixed) also display much higher spread in their orbital energy values at lower $\tau_{\mathrm{infall}}$ values. This is not observed for the lower mass coherent stream/dwarf mergers that are accreted at similar infall times. We speculate that this is due to more massive mergers suffering more from dynamical friction forces than lower mass accretion events. Dynamical friction is the culprit for the stripping of the orbital energy and angular momentum of these more massive disrupted satellites, forcing them to sink deeper into the host's gravitational potential. This will be further investigated in an upcoming study (Donlon et al., in prep). Conversely, at $\tau_{\mathrm{infall}}$ $>$ 10 Gyr, we see that mergers of all masses display a wide range in the spread values of their orbital energies.

\item There is no direct relation between stellar mass and infall time with a disrupted satellite's mean orbital circularity value, suggestive that these two parameters should not play a major role in the radialisation of a debris from a cannibalised satellite system. A detailed study of the time evolution of orbital circularity for different mass disrupted satellites is subject of ongoing work, see Donlon et al (in prep).

 \item On average, coherent stream and bound dwarf accretions present higher mean circularity values ($\langle$$\eta$$\rangle$ $\sim$0.6-1) when compared to their phase mixed counterparts (that present an average value of $\langle$$\eta$$\rangle$ $\sim$0.2-0.4). However, we do note some exceptions at very high $\tau_{\mathrm{infall}}$, where some phase mixed mergers reveal a circularity value close to $\langle$$\eta$$\rangle$ $\sim$0.8. Furthermore, at lower $\tau_{\mathrm{infall}}$ we find that coherent streams can present lower circularity values, reaching a value of $\langle$$\eta$$\rangle$ $\sim$0.2.
 
 \subsubsection{Chemical properties}
 
 \item The distribution of the accretion events in the [Mg/Fe]-[Fe/H] plane reveals that these chemical properties are clearly correlated with an accretion event's stellar mass and time of accretion, given the iso-mass and iso-age trends observed in the left and right panels of Fig~\ref{mgfe}, respectively. Here, more massive accretion events (typically classified as phase mixed) present higher mean [Mg/Fe] for fixed metallicity when compared to their lower mass mergers, which are typically classified as coherent streams/dwarfs. 
 
 \item We find that these more massive accretion events that present higher mean [Mg/Fe] values are also accreted earlier when compared to the lower mass coherent stream/dwarf mergers that display lower [Mg/Fe] for fixed [Fe/H]. When examining the distribution at fixed [Mg/Fe], we find that the relationship between mass and [Mg/Fe] (at fixed [Fe/H]) evolves with $\tau_{\mathrm{infall}}$, which we attribute to the evolution of the mass-metallicity relation of these galaxies across time. Interestingly, our results reveal that on average dwarf galaxies have lower mean [Mg/Fe] (across a range in [Fe/H]) when compared to mergers classified as phase mixed/coherent streams (consistent with the results from \citet[][]{Cunningham2022}).
\end{itemize}

\subsection{Do expectations from prior simulations survive in a high-resolution cosmological setting?}

Overall, our results show that, while the stellar halos of galaxies and the process of hierarchical mass assembly are extremely convoluted, the distribution of the average values of the observable properties of the debris from disrupted satellites follow general expectations from prior simulations when examining them in a high-resolution cosmological setting.

In terms of their spatial and orbital properties, we have found that accretion events that are either more massive, or are consumed by the larger host at an earlier time, deposit the majority of their stars deeper in the potential well of their host, at lower orbital energies and smaller galactocentric radii values. This result corroborates previous work using $N$-body simulations (\citealp[e.g.,][]{Bullock2005,Johnston2008,Wetzel2011,Amorisco2017,Jean2017}), and large-volume cosmological simulations with lower resolution \citep[e.g.,][]{Cooper2010,Pillepich2014,Dsouza2018,Pfeffer2020}. It is also in line with recent estimates from cosmological simulations of similar resolution (\citealp[e.g.,][]{Fattahi2020,Font2020,Grand2021,Khoperskov2022_2,Orkney2022}).

Regarding the chemical compositions of disrupted satellites, our findings are also in qualitative agreement with both results from N-body models \citep[e.g.,][]{Font2006}, large volume cosmological simulations at lower resolution (\citealp[e.g.,][]{Mackereth2019}), and zoom-in simulations of similar resolution \citep[e.g.,][]{Khoperskov2022}.

The advent of the new development of the FIRE-3 simulations \citep[][]{Hopkins2022} which will provide updated stellar evolution models, chemical evolution yields, microphysics and updated fitting functions, as well as the advance of other sophisticated cosmological numerical simulations (e.g., ARTEMIS \citealp[][]{Font2020}, COLIBRE) will pave the way for a more detailed study of the chemo-dynamical properties of stellar halos in galaxies like our own MW at higher resolution and/or with larger statistical samples. Of particular interest for this topic will be the FIRE-3 simulations, and their ability to implement different nucleosynthetic yields post-processing. This, in combination with exquisite observational data from current/upcoming surveys will enable a full characterisation of the stellar halo of the MW, and will allow for a full unveiling of the Galaxy's accretion and mass assembly history.

\subsection{What do these results mean for future surveys?}

\label{implications}
The findings presented in this work have several observable implications for the stellar halo of the Galaxy. The fact that (on average) disrupted satellite galaxies of larger stellar mass deposit the bulk of their stars closer to the centre of the MW-like host suggests that the inner regions of the stellar halos of MW-like galaxies are dominated by one to two massive accretions \citep[][]{Dsouza2018}. This prediction is in-line with recent observational discoveries of accreted stellar populations within the inner regions of the Galactic stellar halo conjectured to be the debris from massive accreted satellites (\citealp[][]{Belokurov2018,Helmi2018, Horta2021}). However, interestingly our results also suggest that a significant number of smaller mass satellites accreted at very early times also must inhabit this region of the MW's stellar halo. Recent observational work has also suggested this is the case for the MW (\citealp[e.g.,][]{Myeong2019,Koppelman2020,Naidu2020}). However, the nature of \textit{all} these recently identified stellar halo populations stills needs to be fully established, as many of the identified halo substructures so far could be the result of one overall ancient massive merger (see the results from \citet[][]{Horta2022} for an example). As we have seen in Fig~\ref{rgc_tinfall}, disrupted satellites can deposit their stars in a wide range of galactocentric radii, which can also appear fragmented in orbital space \citep[][]{Jean2017, Koppelman2020, Naidu2021}. The advent of upcoming stellar surveys such as WEAVE, 4MOST, SDSS-V, and later data releases of the \textit{Gaia} mission will hopefully supply the necessary chemo-dynamical information to fully characterise these stellar populations, which in turn should help settle this matter.

In a similar fashion, our results also suggest that lower mass --some of which are also early accreted-- disrupted satellites should be dominant at 
at larger galactocentric distances. The debris of such mergers should have also remained coherent in phase space and occupy large galactocentric distances (see for example Fig~\ref{rgc_tinfall} and Fig~\ref{orbits_tinfall}), making them detectable in kinematic samples. For the case of these early accreted satellites, this provides a window to study on a star-by-star basis the debris from galaxies formed early in the Universe, providing a window to study near-field cosmology. Surveys such as Rubin/LSST, DESI, and (to some extent) \textit{Gaia} will be extremely helpful for probing these outermost regions of the Galactic stellar halo, and will hopefully supply the necessary data to study these relics of early galaxy formation.

Lastly, we note that the high spread in R$_{\mathrm{GC}}$ values suggests that there is likely a wealth of halo substructure in the Galactic stellar halo that is yet to be characterised and that belongs to accretion events already discovered. Characterising the properties of these stellar populations, and comparing them to theoretical models, will lead to a deeper understanding of how satellite galaxies get engulfed by larger mass systems, helping place constraints on the Galaxy's hierarchical mass assembly.

\section*{Acknowledgements}
DH thanks the Melissa Ness for helpful discussions. He also thanks Sue, Alex, and Debra for their constant support. ECC acknolwedges support for this work provided by NASA through the NASA Hubble Fellowship Program grant HST-HF2-51502.001-A awarded by the Space Telescope Science Institute, which is operated by the Association of Universities for Research in Astronomy, Inc., for NASA, under contract NAS5-26555. RES acknowledges support from NSF grant AST-2007232 and NASA grant 19-ATP19-0068, from the Research Corporation through the Scialog Fellows program on Time Domain Astronomy, and from HST-AR-15809 from the Space Telescope Science Institute (STScI), which is operated by AURA, Inc., under NASA contract NAS5-26555. AW received support from: NSF via CAREER award AST-2045928 and grant AST-2107772; NASA ATP grant 80NSSC20K0513; HST grants AR-15809, GO-15902, GO-16273 from STScI. CAFG was supported by NSF through grants AST-1715216, AST-2108230,  and CAREER award AST-1652522; by NASA through grants 17-ATP17-0067 and 21-ATP21-0036; by STScI through grants HST-AR-16124.001-A and HST-GO-16730.016-A; by CXO through grant TM2-23005X; and by the Research Corporation for Science Advancement through a Cottrell Scholar Award.

The simulations were ran using: XSEDE, supported by NSF grant ACI-1548562; Blue Waters, supported by the NSF; Frontera allocations AST21010 and AST20016, supported by the NSF and TACC; Pleiades, via the NASA HEC program through the NAS Division at Ames Research Center.

\bibliography{refs}

\end{document}